\documentclass[%
 reprint,
superscriptaddress,
amsmath,amssymb,
aps,
pre,
showpacs,
longbibliography
]{revtex4-1}

\usepackage[utf8]{inputenc}
\usepackage{graphicx}
\usepackage{dcolumn}
\usepackage{bm}
\usepackage{tikz}
\usepackage{tabularx}
\usepackage{amsmath}

\usepackage{amssymb}
\usepackage{pgfplots}
\pgfplotsset{compat=1.3}
\usetikzlibrary{plotmarks}
\usepackage{rotating}
\usepackage{hyperref}

\begin{document}

\title{Synthetic Schlieren -- application to the visualization and characterization of air convection}

\author{Nicolas Taberlet}
 \affiliation{Univ Lyon, ENS de Lyon, Univ Claude Bernard Lyon 1, CNRS, Laboratoire de Physique, F-69342 Lyon, France}

\author{Nicolas Plihon}
 \affiliation{Univ Lyon, ENS de Lyon, Univ Claude Bernard Lyon 1, CNRS, Laboratoire de Physique, F-69342 Lyon, France}
 
\author{Lucile Auzémery}
 \affiliation{Univ Lyon, ENS de Lyon, Univ Claude Bernard Lyon 1, CNRS, Laboratoire de Physique, F-69342 Lyon, France} 
 
\author{Jérémy Sautel}
 \affiliation{Univ Lyon, ENS de Lyon, Univ Claude Bernard Lyon 1, CNRS, Laboratoire de Physique, F-69342 Lyon, France}

\author{Grégoire Panel}
 \affiliation{Univ Lyon, ENS de Lyon, Univ Claude Bernard Lyon 1, CNRS, Laboratoire de Physique, F-69342 Lyon, France}

\author{Thomas Gibaud}
 \email{Corresponding author, thomas.gibaud@ens-lyon.fr}
 \affiliation{Univ Lyon, ENS de Lyon, Univ Claude Bernard Lyon 1, CNRS, Laboratoire de Physique, F-69342 Lyon, France}%

\date{\today}

\begin{abstract}

Synthetic schlieren is an digital image processing optical method relying on the variation of optical index to visualize the flow of a transparent fluid. In this article, we present a step-by step, easy-to-implement and affordable experimental realization of this technique. The method is applied to air convection caused by a warm surface. We show that the velocity of rising convection plumes can be linked to the temperature of the warm surface and propose a simple physical argument to explain this dependence. 
Moreover, using this method, one can reveal the tenuous convection plumes rising from ounce’s hand, a phenomenon invisible to the naked eye. This spectacular result may help student realize the power of careful data acquisition combined with astute image processing techniques.

\end{abstract}

\pacs{xxx}
                             
\maketitle

\section{Introduction}

Optical methods such as shadowgraphy \cite{de2011, rasenat1989, dvovrak1880}, Mach–Zehnder interferometry \cite{mach1879} or Schlieren methods \cite{toepler1906,merzkirch2012} provide a dynamic and non-intrusive method to visualize small variations in the optical index of refraction $n$ of a transparent media \cite{settles2012}. The schlieren technique is particularly spectacular and simple/low cost to implement and as such make a good candidate for student lab classes. Schlieren experiments are widely used in fluid dynamics to study physical phenomena where the index of refraction of the media is affected, like in shock waves \cite{clarke2007, pandya2003, pulkkinen2017}, heat emanating from a system \cite{lewis1987,alvarez2009,prevosto2010} or internal waves \cite{sutherland1999, bourget2013}.

The physical basis for schlieren imaging emerges from geometrical optics principles. In an homogeneous transparent media the light rays propagate uniformly at a constant velocity. However, in the presence of spatial  variations of the index of refraction, light rays are refracted and deflected from their continuous path according to Snell’s Law of refraction. Schlieren experiments take advantage of the rays deflection to create a contrasted images that map the variations of the index of refraction.

The name Schlieren experiments regroups a large variety of setups \cite{settles2012, settles2017}. 
The first use of the Schlieren technique dates back to the end of the XIXth century by Toepler and led, for instance, to the first observation of shock waves. It is based on imaging the variations of the index of refraction using a knife edge blocking part of the light rays (which can be understood as a Fourier optical filtering system)~\cite{gopal2008}. Here we focus on 'synthetic schlieren' a variation of the Schlieren methods which was originally developed by Sutherland, Dalziel, Hughes and Linden in 1999 \cite{dalziel2000, sutherland1999} and applied for instance to natural convection problems~\cite{Ambrosini2006}. This method relies on imaging a patterned through a  media with varying index of refraction. Digital image processing then allow to infer the variation of the index of refraction and to relate them to variations of the physical properties of the media. 
The experiment and the image analysis leading to the map of the variations of the index of refraction of a slab of fluid are descibed in section~\ref{sec:ns}. This technique is then applied to the visualization and characterization of air convection induced by a heat source in section~\ref{sec:convection}. We show in particular that it is possible to visualize and measure the heat of hands being rubbed together.

The pedagogical interest of the synthetic schlieren method is manifold. Not only is it a direct illustration of the principles of geometrical optics and thermal convection, but it also illustrates the remarkable efficiency of interferometric techniques. Indeed, the method makes use of slight differences (whose typical size is only a fraction of the probing wavelength) in images to reveal fine details of an otherwise-invisible phenomena (as demonstrated by the convection plumes rising from a hand, see Fig.~\ref{fig:hand}). Moreover, this paper constitutes an interesting introduction to data analysis and image processing which are now widely used in undergraduate lab projects thanks to the spread of affordable high-speed digital cameras. Using a user-friendly free software, one is able to visualize a seemingly invisible flow in a few simple steps, and can extract the velocity of rising convection plumes using an insightful built-in space-time tool.

\section{Synthetic schlieren experiment}
\label{sec:ns}

\subsection{Experimental setup}
\label{sec:setup}

\begin{figure}
	\centering
  \includegraphics{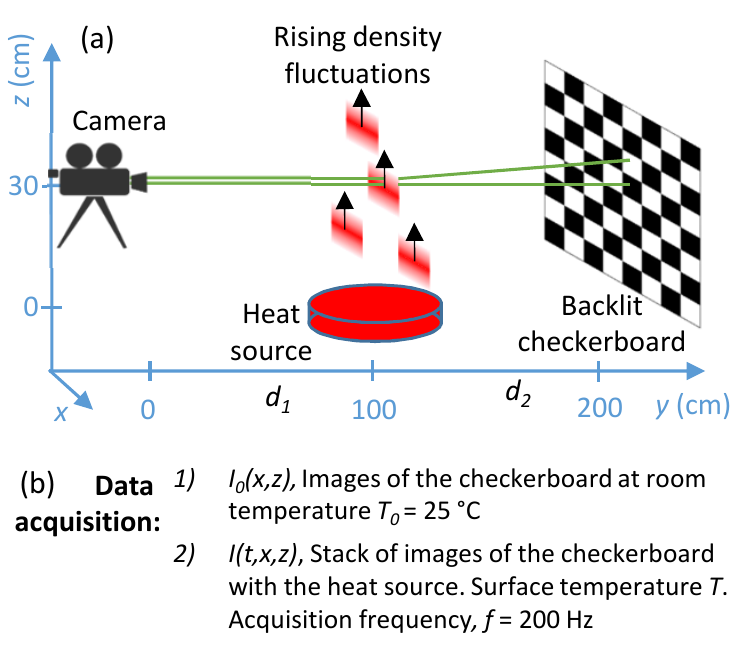}
     \caption{Schematic of the synthetic schlieren experiment. A camera is focused on a checkerboard printed on a transparent plastic sheet and back illuminated by a led panel. A heat source is placed between the camera and the checkerboard and induces air convection. The air convection is associated to a temperature, density and index of refraction gradient in the air just of above the heat source surface. The cherkerboard image is altered because the light rays are deviated toward the high index of refractions. The analysis of this alteration induced by variation of the index of refraction is at the core of the synthetic schlieren experiment. (b) Data acquisition procedure.
}
    \label{fig:setup}
\end{figure}

The experimental setup to perform synthetic schlieren is shown in Fig.~\ref{fig:setup}. A camera (Ximea xiQ model MQ013MG-ON) with an objective with a zoom 12.5-75~mm is mounted on tripod at $y=0$~cm and $z=30$~cm. The camera focuses on a checkerboard at $y=d_1+d_2=200$~cm. We use the zoom to image the full checkerboard. The checkerboard is printed on a transparent plastic A4 sheet and composed of black squares and transparent squares of equal dimensions $L=0.9$~mm. Backlighting of the sheet is provided by a led panel to allow for fast acquisition rates, $f$, typically 25 to 200 Hz. 

The experiment takes place in two steps. First, a reference image $I_0(x,z)$ of the checkerboard is acquired. Then a physical phenomena that disturb the air index of refraction in the camera field at $y=d_1=100$~cm between the camera and the checkerboard is turned on. In this article, we illustrate the synthetic schlieren technique using the air convection produced by a heat source. The air convection is associated in the $z$-direction to a gradient of temperature $\nabla T$, density $\nabla \rho$ and index of refraction $\nabla n$ of the air just of above the heat source surface. The checkerboard image $I(t,x,z)$ is recorded as function of the time $t$ at a rate $f$ by the camera. 
As light rays are deviated toward the higher values index of refractions, the index of refraction variations $\nabla n$ modify the image of the checkerboard $I$ as to compared to the reference image $I_0$. Visualisation of the volutes is then achieved from the analysis of this alteration. The heat source temperature may also be determined, provided a calibration procedure.

$I$ is changed compared to $I_0$. Indeed, due to the $\nabla n$, the light rays are deviated toward the high index of refractions. The analysis of this alteration permits to visualize the volute and even determined the heat source temperature provided the experiment is calibrated. 

\subsection{Data analysis}
\label{sec:data}

\begin{figure}
	\centering
  \includegraphics{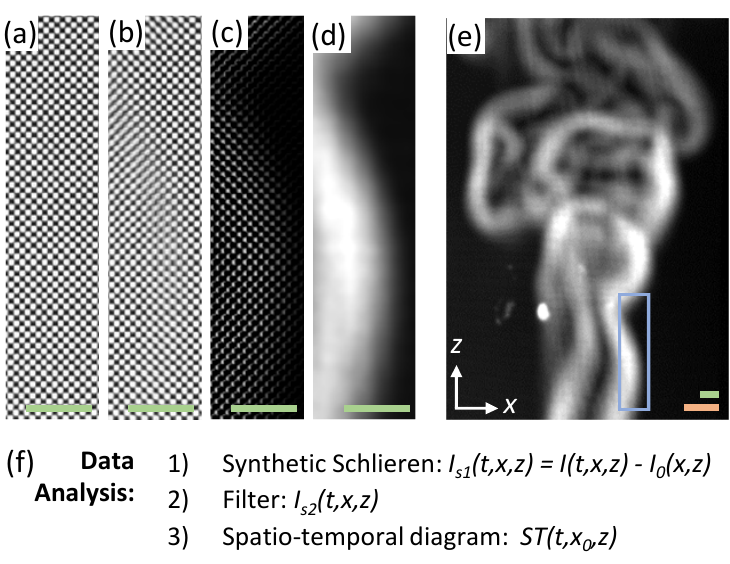}
     \caption{Typical data processing. The heat source is the flame from a lighter. (a) Image of the checkerboard $I_0(x,z)$ at room temperature $T_0=$25~$^\circ$C. (b) Image of grid $I(t,x,z)$ when the hit source is on. (c) Image resulting from the synthetic schlieren, $I_{s1}(t,x,z)$.  (d) Filtered Image, $I_{s2}(t,x,z)$. (e) overall image $I_{s2}$. The blue box indicates the region of interest display in (a-d). There are two scale bars: the green upper one is related to the grip plan and the lower orange one is related to the heat source plan. Both represent 1~cm. (f) Analysis procedure
}
    \label{fig:ana}
\end{figure}

The data reduction used in this article is a three-step process that can be done with ImageJ a free software dedicated to image analysis \cite{imagej}. Those steps are described in the supplementary materials \cite{SM}. 
For each image, the first two steps provide an image displaying the intensity of the variation of the index of refraction. This series of images can then be either analysed as a movie or, following the approach described as a third step in this article, using spatio-temporal diagrams. 

The first two steps are illustrated in Fig.~\ref{fig:ana}. $I_0$ is subtracted to each image $I(t,x,z)$ to provide a stack of images $I_{s1}(t,x,z)$. When there is no index of refraction variations, the image $I(t,x,z)$ is identical to $I_0$ and difference is null, leading to a black $I_{s1}$ image. In the presence of variations of the index of refraction, $I(t,x,z)$ is a distorted image of the checkerboard and he resulting image $I_{s1}(t,x,z)$ shows grey spots which spatial extension $\delta r$ is related to the intensity of the fluctuations of  refractive index in the $y=d_1$ plan. These grey spots in $I_{s1}$ are then low pass filtered to form a continuum $I_{s2}$ image. Either a low-pass or a Gaussian blur filter are used with a cutoff length $l_c$ of at least twice the checkerboard periodicity, ($l_c= 4L=3.6$~mm). Fig. \ref{fig:ana}(e) shows $I_{s2}$ the resulting schlieren image induced by the flame of a lighter. We observe volutes rising over time. $I_{s2}$ display two scale bars. The green one corresponds to the scale bar in the checkerboard plan. The orange one corresponds to the scale of objects placed in the heat source plane. Due to the setup geometry there is a perspective effect and objects position at $y=d_1$ appear lager than objects placed at $y=d_1+d_2$.

\begin{figure}
	\centering
  \includegraphics{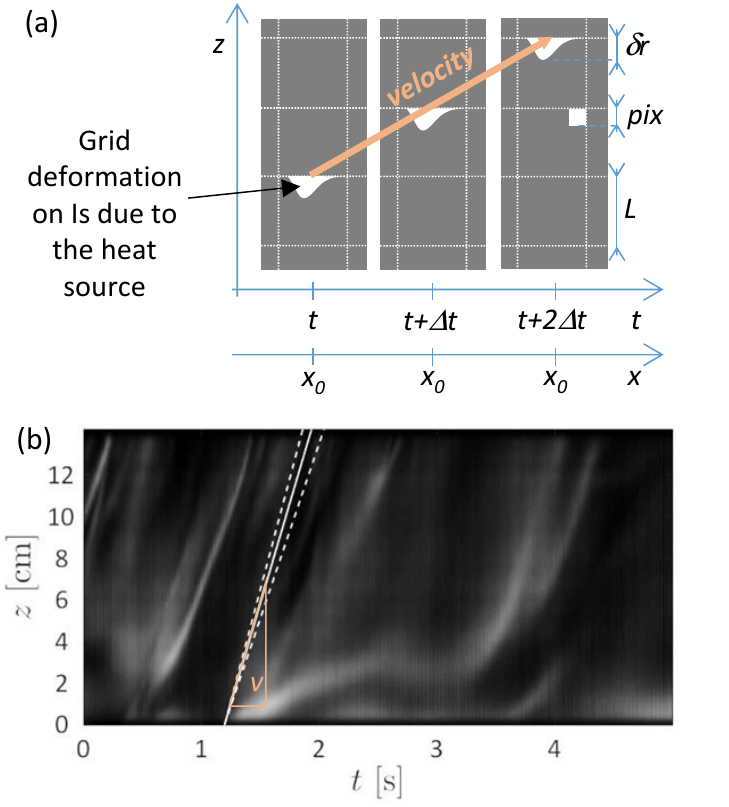}
     \caption{Spatio temporal diagram. (a) The synthetic schlieren experiment produces a stack of images  $I_{s2}(t,x,z)$ that display the local checkerboard deformation of spatial amplitude $\delta r$. The setup is chosen so that $\delta r$ is about the pixel size but smaller than the grid size $L$. For a hot heat source those deformations propagate upward. Following those deformation permits to determine the rising velocity $v$ of the volute. (b) $ST(t,x_0,z)$ is the spatio temporal diagram of the stack $I_{s2}$. $ST$ is obtained by choosing a horizontal position $x_0$ and plotting $I_{s2}(t,x_0,z)$ as a function of $t$ and $z$. The $z$ axis is graduated following the scale bar related to the heat source plan. The inclination of the pattern indicates the volute velocity $v$. An aluminum block at $T=81$~$^\circ$C yield $v= 70 \pm 8$~cm/s 
}
    \label{fig:st}
\end{figure}

In this article, we detail a last data reduction step permits to visualize the dynamics of the volutes. It consists in computing the spatio temporal diagram $ST(t,x_0,z)$ of the stack of images $I_{s2}$, as displayed in Fig.~\ref{fig:st}. $ST$ is obtained by choosing a horizontal position $x_0$ and plotting $I_{s2}(t,x_0,z)$ as a function of $t$ and $z$. The $z$ axis is graduated following the orange scale bar related to the heat source plan. In this representation we observe grey inclined lines which are the trace of the rising volutes. Around 5~cm above the heat source, the line inclination becomes constant over time. In this region their inclination yields a stationary volute rising velocity $v$. By measuring different line inclinations within the time frame of the experiment, typically a few seconds, and at different $x_0$ we determine the average velocity $v$ of the volute and its standard deviation $dv$. We typically find that $dv/v\sim 20$\%.

\subsection{Setup parameters}
\label{sec:para}

Obviously $I_0$ and $I$ must be acquired in the same conditions. If the checkerboard has moved from one acquisition to the other, a moiré pattern is observed in the background of $I_{s1}$.

This experiment depends crucially on a few parameters: $d_1$ the distance between the camera and the heat source, $d_2$ the distance between the heat source and the checker board, $L$ the square size on the checkerboard, and $f$ the acquisition frequency. 

The value of $d_1$ and $d_2$ set the orange and green scale bars on $I_{s2}$. Small $d_1$ and large $d_2$ enlarge the orange scale compared to the green scale. It follows that the technique sensitivity to $\nabla n$ is increased. We choose $d_1=d_2$ so that their is roughly a factor two between the orange and green scale bar.

The length $L$ also set upper bound on the variations of index of refraction that can be unambiguously determined.
Denoting $\delta r$ the spatial amplitude of the checkerboard deformation, a given value of $L$ is limited to measure deformation such that 1~$pix<\delta r<L$. If $\delta r< 1$~$pix$, $\nabla n$ is too small and no deformation is measured. If $\delta r>L$, the local image deformation is larger than the checkerboard dimension and leads to aliasing and to erroneous volute velocity measurements.

Finally, the length $L$ must not be a multiple of a pixel or moiré effect are observed \cite{marsh1980, amidror2000}.

The low-pass filter characteristic length defines the spatial resolution with which we observe the volute on $I_{s2}$. In the present article, it has been chosen as $l_c=2L$, but larger values are possible.

Last, the acquisition frequency $f$ must be chosen accordingly to the velocity of the phenomena studied. The typical time scale can either be derived from physical modeling, or, practically from the analysis of the spatio-temporal diagrams. In the case of convection, if $f$ is too high, the inclination of the grey lines in the spatio-temporal diagram are almost vertical and this gives a poor resolution on the velocity $v$. If $f$ is too low, the volutes travel too much of a distance between two frames and the grey inclined lines disappear, it becomes impossible to measure $v$.

\section{Application to the air convection induced by a heat source}
\label{sec:convection}

To benchmark the schlieren experiment, we choose to study the air convection induced by a heat source \cite{batchelor1954}. This is typically the situation of rising smoke from a fire. In this case the dust permits to visualize the volutes. However, in the absence of dust, volutes cannot be seen by the naked eyes. Here, as seen in previous section the schlieren experiments permits to visualize the volutes and measure their rising speed. An other common example of natural convection are hot mirages \cite{ vollmer2003, zhou2011, berry2013}. In a hot mirage the sun heats the ground and creates a time averaged temperature gradient $\langle\nabla T\rangle$ between the ground surface at temperature $T$ and the surrounding air at temperature $T_0<T$ in a boundary layer. The temperature gradient goes along with a refractive index gradient $\langle\nabla n\rangle$: moving away from the ground $\langle n\rangle$ increases. The light rays coming from above the horizon are bend toward high $n$, reach the eyes of an observer and give him the impression that the sky is on the ground. Superposed to this time-averaged description, the hotter and lighter air close to the ground dynamically rises up. These rising volutes dynamically blur the scenery. Our digital schlieren setup is calibrated from the dynamical analysis of the rising volutes induced by a surface of know temperature $T$.

\subsection{Calibration}
\label{sec:cal}

\begin{figure}
	\centering
  \includegraphics{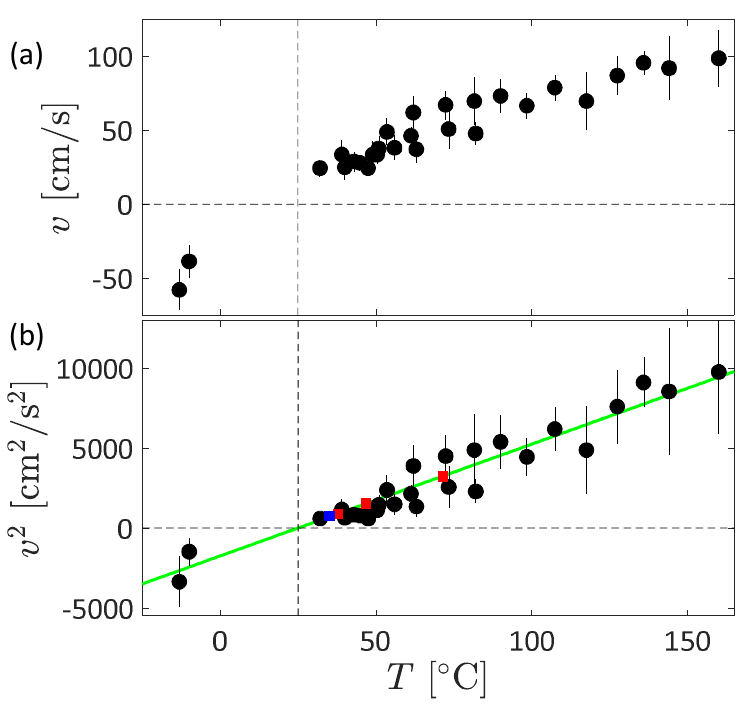}
     \caption{Volute velocity $v$ with its corresponding standard deviation $dv$ as a function of the heat source temperature $T$. (a) $v$ as a function of T. (b) $v^2$ and its direction as a function of $T$. $circ$ corresponds to the calibration experiments and are obtained by measuring the volute velocity induced by a aluminum block with a surface temperature $T$. $T$ is measured independently by a thermometer embedded in the block. The line corresponds to the best linear fit, $v^2=A(T-T_0)$ with $A=65\pm 6$~cm$^2$s$^{-2}$K$^{-1}$ and $T_0=25$~$^\circ$C. Blue and red Square correspond the volute velocities induced by the surface of a hand and a rubbed hands respectively. In this case, $v$ is measured and positioned on the fit line so that one can determine the hand surface temperature using the calibration. We find 35~$^{\circ}$C for the unrubbed hand surface and temperatures up to 73~$^{\circ}$C for the rubbed hand surface (see Fig.~\ref{fig:temp}).
}
    \label{fig:temp}
\end{figure}

First we measured the velocity of volute $v$ of volutes created by the air convection triggered by the presence of a hot aluminum block (radius 6.1~cm, height 3~cm). The aluminum block was heated up in an oven at 200~$^\circ$C. A small hole in the aluminum block was drilled so that we could monitor its temperature $T$ with thermocouple thermometer. We then placed the aluminum block in the schlieren experiment and performed images acquisitions as its temperature slowly decreases to $T_0$. A similar experiment was conducted using an aluminum block cooled down in a freezer; in this case the volutes move downward. Fig.~\ref{fig:temp}(a) shows the evolution of $v$ as a function of $T$. We observe that the velocity increases monotonously with $T$. The volute velocity is characteristic of the block temperature. Fig.~\ref{fig:temp}(b) shows that $sign(v)v^2$ increases linearly with temperature. A linear fitting procedure provides a temperature calibration of the surface of the block. This calibration indeed depends on the details of the experimental setup.

\subsection{Surface temperature of rubbed hands}
\label{sec:hand}

\begin{figure}
	\centering
  \includegraphics{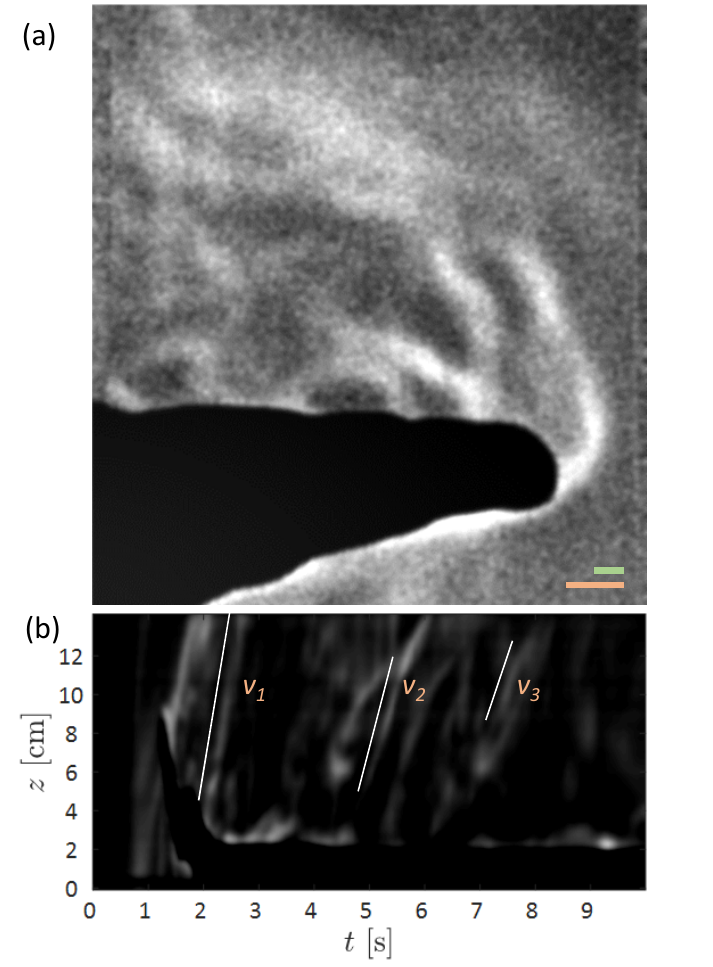}
     \caption{Surface temperature of a rubbed hand using synthetic schieren. (a) Synthetic schlieren results of a rubed hand. the green upper one is related to the grip plan and the lower orange one is related to the heat source plan. Both represent 1 cm. (b) Spatio temporal diagram of image stack corresponding to (a). The volutes speed decreases with time ($v_1=57$ $v_2=43$ and $v_3=30$~cm/s) which corresponds to decreasing temperatures ($T_1=73$, $T_2=53$ and $T_3=38$~$^{\circ}$C).
}
    \label{fig:hand}
\end{figure}


Having set up a calibration, we are now in a position to determine any surface temperature between -10 and 150~$^\circ$C using the synthetic schlieren setup in Fig.~\ref{fig:setup}. We choose to look at the temperature of hands rubbed against one another. Just after being rubbed, one of the hand is place in the experimental setup. The hand surface temperature is hight enough with respect to $T_0$ to observe volutes, as diplayed Fig.~\ref{fig:hand}(a). The spatio temporal diagram in Fig. \ref{fig:hand}(b) shows that the hand temperature decreases from 73~$^\circ$C to its equilibrium temperature, 35~$^{\circ}$C, within $\sim 10$~s.

\subsection{Discussion}
\label{sec:disc}

In this subsection, we derive a simple model for the estimate of $v$ and for the scaling $v^2 \sim T$.

\begin{figure}
	\centering
  \includegraphics{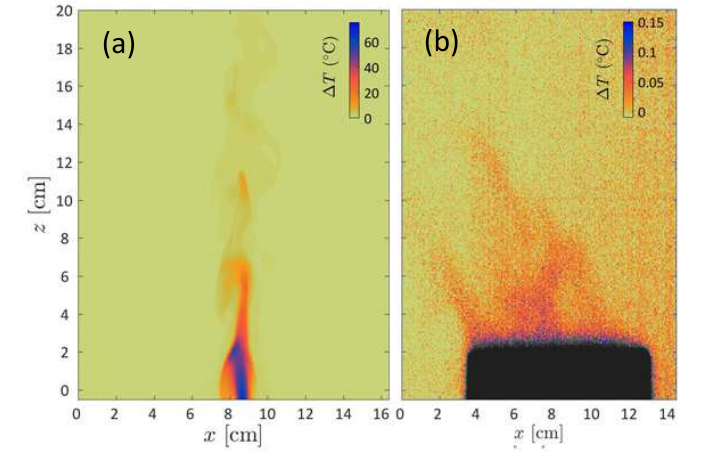}
     \caption{Images normalized by the background temperature obtained from an infrared camera. The Normalization is obtained by subtracting to the image with the hot source the image at room temperature. (a) Flame at $T=900$~$^{\circ}$C. (b) Aluminum block at $T=66$~$^{\circ}$C. The block intensity is saturated and temperature scale does not apply on the block.
     }
    \label{fig:therm}
\end{figure}

To model the volute speed $v$, it is necessary to estimate or measure the air temperature within the volute. The volute temperature is indeed responsible for the volute lower density and therefore its rise. To do so we used an infrared camera (Flir) which acquire images where the intensity codes for the temperature. Fig.~\ref{fig:therm} shows the volutes temperature measured by an infrared camera for a (a) flame and (b) the aluminum block experiment described previously (note the difference in temperature scale between both images) . The volute temperature is much lower than the source temperature. For a flame, the volute temperature is $\Delta T_{air} \sim 30$~$^{\circ}$C above room temperature. For an aluminum block at $T=66$~$^{\circ}$C the volutes temperature are barely measurable with the infrared camera, $\Delta T_{air} \sim 0.15$~$^{\circ}$C. In this context, the affordable synthetic schlieren setup proposed here proves to be more sensitive and appropriate to visualize hot air volutes than expensive infrared cameras.

We have now all the ingredients necessary to estimate with a simple model the volute velocity. Based on the schlieren images $I_{s2}$, we assume that the volute is shaped like a streamed body of radius $r=1$~cm and hight $h=10$~cm (Volume, $V_{sb}\sim \frac{4}{6}\pi r^3+\frac{1}{3}\pi r^2 h$). This is the first strong hypothesis; the volute is indeed not a solid body as the hot air circulate within the volute. The volute is subject to the buoyancy force $\Pi_a$ and the drag force $f_v$. $\Pi_a$ is due to the mismatch of density between the volute and surrounding air and it drive the volute to rise against the gravity field $g$. $f_v$ is due to the air resistance and is opposed to the rising motion of volute. The Reynolds number having a value of $Re=vh/ \nu\sim 3000$ (air dynamic viscosity $\nu=15$~$\mu$m$^2$/s) the drag coefficient for a stream body is $C=0.04$. This value is effective for a solid/fluid interface and therefore constitute the second strong hypothesis; the volute interface is indeed fluid/fluid. In the stationary regime, where the volute velocity is constant, the two forces compensate each other and we obtain the following expression for $v$:

\begin{equation*} \label{eq1}
	\left\{
	\begin{array}{rl}
\Pi_a & =  V_{sb} \Delta \rho g \\
 & \\
f_v & =  \frac{1}{2}C \rho \pi r^2 v^2
	\end{array}
	\right.
\end{equation*}
\begin{equation} \label{eq1}
   \Rightarrow v^2  \sim \frac{2V_{sb}g}{C \pi r^2} \frac{\Delta \rho}{\rho}.
\end{equation}

Using the ideal gas law for the air and differentiating it at constant pressure $P$ and particle number $N$, we obtain a relation between  $\Delta T_{air}$ and $\Delta \rho$:

\begin{equation} \label{eq2}
P  = \frac{N}{V}RT_{air} 
\Rightarrow \frac{\Delta \rho}{\rho}=-\frac{\Delta T_{air}}{T_0}.
\end{equation}

Assuming that the pressure remains constant constitute the third strong hypothesis in this model. Using Eq. \ref{eq1} and \ref{eq2}, we obtain an expression for the volute velocity as function of $\Delta T_{air}$:

\begin{equation} \label{eq3}
v \sim \sqrt{ \frac{2V_{sb}g}{C \pi r^2} \left|\frac{\Delta T_{air}}{T_0}\right|}.
\end{equation}

For the aluminum block at $T=66$~$^{\circ}$C, the infra red camera yields a volute temperature difference $\Delta T_{air}=0.15$~$^{\circ}$C with respect to the room temperature $T_0=25$~$^{\circ}$C. Using Eq. \ref{eq3}, we find $v\sim 10$~cm/s. This is give the right order of magnitude. The model however underestimate the experimental value of 50~cm/s.

This simple model yields that $v^2 \sim \Delta T_{air}$. Provided that the temperature of the volute is proportional to the surface temperature of the heat source, the model explains the scaling observed empirically in Fig.~\ref{fig:temp}(b).

\section{Conclusion}
\label{sec:conc}

There are many optical techniques that use the deflection or  phase  changes in light rays, to map  variation  in  the  refractive index $n$. The classical shadowgraph  method  \cite{de2011, rasenat1989, dvovrak1880} is  sensitive to the curvature in the refractive index field which focuses or defocuses nominally parallel light rays.  This technique is  essentially qualitative because it is difficult to extract quantitative  information about the density fluctuations due to boundary condition at the edges of the field of view. Interferometers such as the Mach–Zehnder interferometer \cite{mach1879}  provides direct measurements of variations in the  speed of light through the phase change experienced by monochromatic light. However, its application is often limited by its cost and the precision required in setting it up. Schlieren methods \cite{toepler1906,merzkirch2012} are sensitive to refractive index  variations in the plane normal to light rays  passing through the medium. While schlieren has been used for many years to visualize flows containing variations in refractive index, its application may be limited by the price of the optical components. Indeed, the visualization of large domains requires the use of expensive parabolic mirrors. It also may be difficulty to extract quantitative information, for instance, in its simplest form, the intensity of a `knife  edge' schlieren image is polluted by a gradients of the refractive index perturbations in the direction of the knife edge. 

Synthetic schlieren is an alternative to those technique. It is simple to setup and cost effective now that fast camera are chip. It is sensitive, fast, local and yields qualitative information about the 2D flow without the use of dyes or tracers. In this article, we first have described the experimental setup, shown how to process the data and discussed the experiment parameters. In a second part, we have successfully tested synthetic schlieren to study the convection induces by a heat source. We have visualized the volute produced by the heat source and measured their velocity $v$. We came up with an empirical scaling that relate unambiguously $v$ to the temperature of the heat source $T$. Building on this calibration, we have fallowed the heat released by rubbed hands. Finally using a simple model we have estimated $v$ and justified the empirical scaling.




\section*{Acknowledgements}
\label{sec:ack}

The authors acknowledge support from the PALSE program of the University of Lyon Saint-Etienne, the University Lyon Claude Bernard, the Soci\'et\'e Fran\c caise de Physique and from the \'Ecole Normale Sup\'erieure de Lyon and its Physics Department and Laboratoire de Physique. We thanks Stéphane Santucci and Kenny Rapina for their help with the infrared camera. The work presented here was done in preparation for the International Physicists Tournament (\url{http://iptnet.info}), a world-wide competition for undergraduate students. The authors are grateful to both local and international organizing committees of the International Physicists Tournament for having put together an exciting event.

\bibliographystyle{apsrev4-1}
\bibliography{biblio}

\end{document}